\documentclass[journal=jacsat,manuscript=article]{achemso}
\usepackage{expl3}
\usepackage{amsfonts}
\usepackage{graphicx, type1cm, lettrine}

\usepackage[version=3]{mhchem} 
\usepackage{color}
\usepackage[colorlinks=true, letterpaper=true, pdfstartview=FitV, linkcolor=blue, citecolor=blue, urlcolor=blue]{hyperref}

\author{Xiaoming Zhang}
\affiliation{State Key Laboratory of Reliability and Intelligence of Electrical Equipment, Hebei University of Technology, Tianjin 300130, China.}
\alsoaffiliation{School of Materials Science and Engineering, Hebei University of Technology, Tianjin 300130, China.}

\author{Botao Fu}
\affiliation{College of Physics and Electronic Engineering, Center for Computational Sciences, Sichuan Normal University , Chengdu 610101, China.}

\author{Lei Jin}
\affiliation{State Key Laboratory of Reliability and Intelligence of Electrical Equipment, Hebei University of Technology, Tianjin 300130, China.}
\alsoaffiliation{School of Materials Science and Engineering, Hebei University of Technology, Tianjin 300130, China.}

\author{Xuefang Dai}
\affiliation{State Key Laboratory of Reliability and Intelligence of Electrical Equipment, Hebei University of Technology, Tianjin 300130, China.}
\alsoaffiliation{School of Materials Science and Engineering, Hebei University of Technology, Tianjin 300130, China.}

\author{Guodong Liu}\email{gdliu1978@126.com}
\affiliation{State Key Laboratory of Reliability and Intelligence of Electrical Equipment, Hebei University of Technology, Tianjin 300130, China.}
\alsoaffiliation{School of Materials Science and Engineering, Hebei University of Technology, Tianjin 300130, China.}

\author{Yugui Yao}\email{ygyao@bit.edu.cn}
\affiliation{Key Lab of advanced optoelectronic quantum architecture and measurement (MOE), and School of Physics, Beijing Institute of Technology, Beijing.}

\title{ Topological Nodal Line Electrides: Realization of Ideal Nodal Line State Nearly Immune from Spin-Orbit Coupling }
\abbreviations{}
\keywords{Electrides, Topological nodal line, First-principles }

\begin{document}

\begin{abstract}
Nodal line semimetals (NLSs) have attracted broad interest in current research. In most of existing NLSs, the intrinsic properties of nodal lines are greatly destroyed because nodal lines usually suffer sizable gaps induced by non-negligible spin-orbit coupling (SOC). In this work, we propose the topological nodal line electrides (TNLEs), which achieve electronic structures of nodal lines and electrides simultaneously, provide new insight on designing excellent NLSs nearly immune from SOC. Since the states near the Fermi level are most contributed by non-nucleus-bounded interstitial electrons, nodal lines in TNLEs manifest extremely small SOC-induced gap even possessing heavy elements. Especially, we propose the family of A$_2$B (A = Ca, Sr, Ba; B= As, Sb, Bi) materials are realistic TNLEs with negligible SOC-induced gaps, which can play as excellent platforms to study the intrinsic properties of TNLEs.
\end{abstract}

\newpage
\section{1. INTRODUCTION}

Topological semimetals have received broad research interest in recent years. To date, several categories of topological semimetals have been well studied, such as Weyl semimetals~\cite{add1,add2,add3,add4,add5}, Dirac semimetals~\cite{add6,add7,add8,add9,add10}, nodal line semimetals (NLSs)~\cite{add11,add12,add13}, nodal surface semimetals~\cite{add14,add15,add16,add17}, and so on. Among them, NLSs have attracted increasing attention in the last couple of years, because they can host many appealing properties, such as exotic drumhead surface states~\cite{add11,add12}, rich transport characters~\cite{add18,add19,add20}, and novel optical responses~\cite{add22,add23}.

\begin{figure}
\includegraphics[width=10cm]{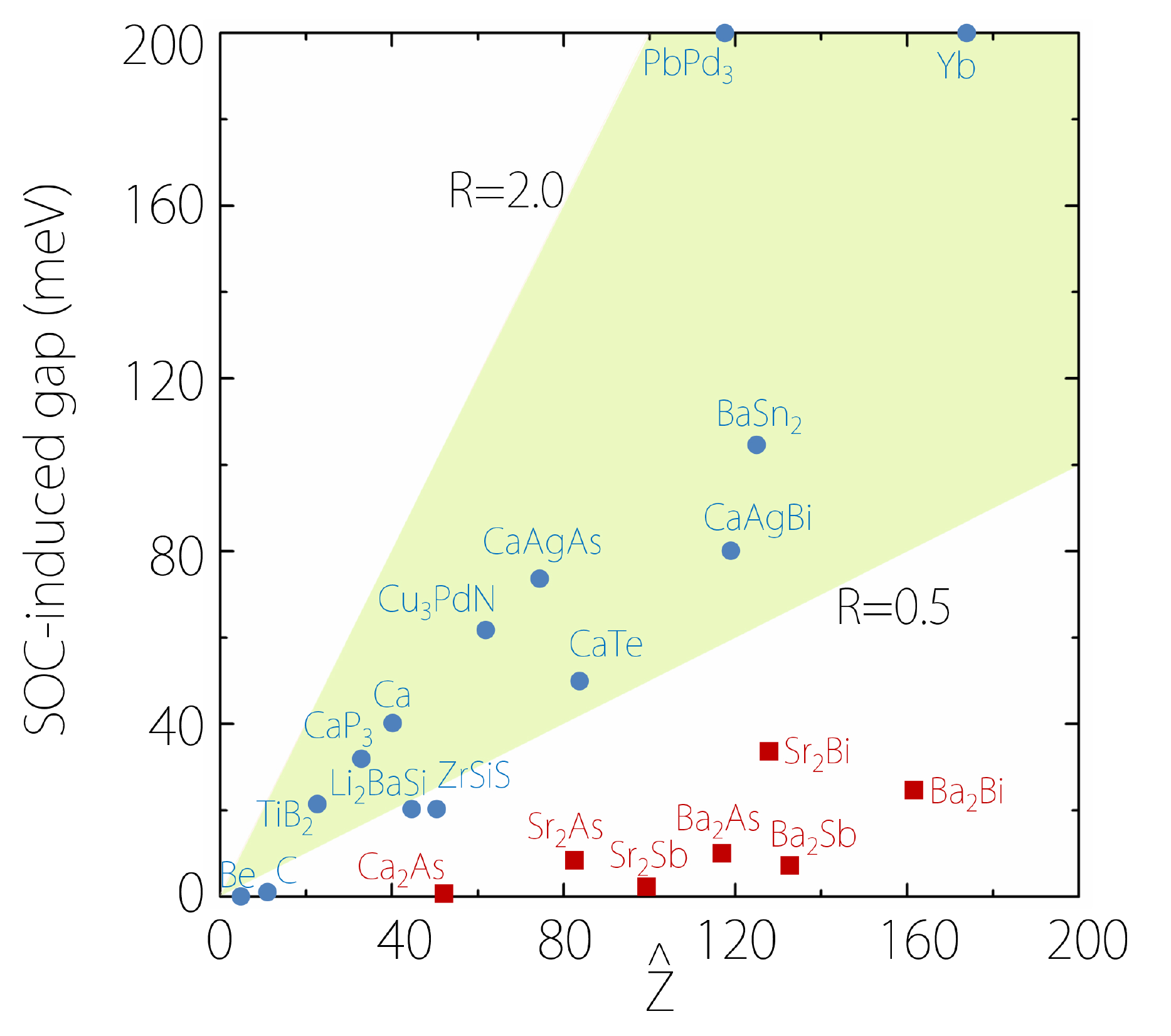}
\caption{The size of SOC-induced gaps around nodal lines as a function of average SOC strength expressed by the average atomic weight ($\hat{Z}$) for present A$_2$B TNLEs and existing typical NLSs. For a given material with A$_{i}$B$_{j}$C$_{k}$ composition, $\hat{Z}$ is calculated as: $\hat{Z}$ = (i*Z$_{A}$+j*Z$_{B}$ +k*Z$_{c}$) / (\emph{i} + \emph{j} + \emph{k}). Note: the size of SOC-induced gaps for A$_2$B TNLEs are represented by the local gap at crossing point A; and that for existing typical NLSs are the local gaps along high-symmetry \emph{k}-paths around the nodal lines, taken from literatures~\cite{add11,add24,add25,add26,add27,add28,add29,add30,add31,add32,add33,add34,add35,add36,add37,add38,add39,add40,add41,add42,add43,add44,add45,add46,add47,add48,add49}. We define the SOC-induced gap ratio R = (gapsize in meV) / $\hat{Z}$, and the shadowed region in the figure is for 0.5 $<$ R $<$ 2.0.
\label{fig1}}
\end{figure}

	Currently, most realistic NLSs are proposed in materials where the inversion (\emph{P}) symmetry and the time reversal (\emph{T}) symmetry coexist, such as graphene networks~\cite{add11,add12}, Cu$_3$PdN~\cite{add24,add25}, CaAgAs~\cite{add26}, AX$_2$ (A = Ca, Sr, Ba; X = Si, Ge, Sn)~\cite{add27}, ZrSiS~\cite{add28,add29,add30}, CaTe~\cite{add31}, some elemental metals~\cite{add32,add33,add34}, Mg$_3$Bi$_2$~\cite{add35}, CaP$_3$~\cite{add36}, MB$_2$ (M = Sc, Ti, V, Zr, Hf, Nb, Ta)~\cite{add37,add38}, Ca$_2$As~\cite{add39}, Li$_2$BaSi materials~\cite{add40}, and so on~\cite{add41,add42,add43,add44,add45,add46,add47,add48,add49}. It is well known that for a spinless system, a local symmetry \emph{P}\emph{T} is enough to protect a nodal line, for which a general effective Hamiltonian may be expressed as~\cite{add11,add2nd1}:
\begin{equation}\label{(1)}
\emph{H$_{0}$}(\emph{k})=(m-k^{2})\tau_{z}s_{0}+k_{z}\tau_{x}s_{0},
\end{equation}
where \emph{k}$^{2}$ = \emph{k}$^{2}_{x}$ + \emph{k}$^{2}_{y}$ + \emph{k}$^{2}_{z}$, \emph{$\tau$$_{0}$$_,$$_{i}$} acts on the orbital index, \emph{s$_{0}$$_,$$_{i}$} acts on the spin space with \emph{$\tau$$_{0}$} and \emph{s$_{0}$} to be the identity matrices. The symmetries can be represented by \emph{P}\emph{T} = \emph{K} with \emph{K} representing complex conjugation operator. After considering the spin-orbital coupling (SOC), the nodal line would be gapped out by symmetry-permitted mass terms. For example, with SOC, the symmetry can be represented as \emph{P}\emph{T}= \emph{i}\emph{s$_y$}\emph{K}, and we can add a mass term

\begin{equation}\label{(2)}
\emph{H}=H_{0} + \Delta\tau_{y}s_{z}
            ,
\end{equation}
with $\Delta$ denoting the effective SOC strength ($\Delta$ is generally relative to the atomic weight, due to the presence of Coulomb potential in the SOC Hamiltonian). One can find that nodal lines are gapped under SOC and the size of gap is relative to $\Delta$. Especially when the gap is sizable, the intrinsic properties of nodal lines are greatly destroyed. An investigation on typical NLSs preserving \emph{P} and \emph{T} symmetries indeed finds the size of gaps is positively related to $\Delta$ expressed by the average atomic weight ($\hat{Z}$), as shown in Fig.~\ref{fig1}. More specifically, by defining the SOC-induced gap ratio [R, expressed as the SOC induced band gap divided by the $\hat{Z}$], it is found that typical NLSs proposed previously almost statistically distribute in a specific region with 0.5 $<$ R $<$ 2.0 (see Fig.~\ref{fig1}). As a result, to obtain excellent NLSs with relatively small SOC-induced gaps, it is used to explore NLSs from materials without containing heavy elements, which however greatly restricts the scale of candidate materials for NLSs. Then, is it possible to develop excellent NLSs in heavy-element-containing materials? One way is to seek spin-orbit-stable nodal line under other protection mechanism (such as the reflection symmetry) rather than \emph{P} and \emph{T} symmetries. However, till now non-centrosymmetric PbTaSe$_2$ is the only experimentally confirmed NLSs robust against SOC~\cite{add50}.

In this work, by combining the electronic properties of electrides, we propose that topological nodal line electrides (TNLEs) can serve as an effective way to realize ideal NLSs with negligible SOC-induced gaps, which can even be applied in heavy-elements containing systems. In electrides, the conduction electrons mostly originate from the so-called excess electrons, being localized in the interstitial sites of the lattice~\cite{add51,add52}. On the one hand, the unique electron states in the electrides are previously proposed to be favorable for obtaining band inversions needed for topological materials~\cite{add57}. As the results, several topological electrides such as Ca$_3$Pb, Y$_2$C, Sc$_2$C, Sr$_2$Bi, LaBr, HfBr, CsO$_3$ are proposed~\cite{add57,add56,add58}. On the other hand, the excess electrons in electrides are not constrained by the nuclei electric field, thereby intrinsically manifest weak SOC effect. Therefore, TNLEs, where the nodal line band structure origins from the non-nucleus-constrained electrons, are expected to obtain ideal NLSs with nearly SOC-free nodal line states even containing heavy elements. Under this fresh viewpoint, in the following we demonstrate the feasibility of realizing TNLEs, and of developing ideal NLSs insensitive to SOC in \emph{P}\emph{T} symmetries preserving system.

\section{2. COMPUTATIONAL METHODS}
We carry out first-principles calculations by using the Vienna ab-initio simulation package based on the density functional theory (DFT)~\cite{new6,new7}. During our calculations, the exchange-correlation potential is chosen as generalized gradient approximation with the Perdew-Burke-Ernzerhof (PBE) realization~\cite{new8}. The nonlocal Heyd-Scuseria-Ernzerhof (HSE06) hybrid functional is also used to check the band structure~\cite{new9}. The cutoff energy during the calculations is chosen as 450 eV. To sample the Brillouin zone (BZ), a 11$\times$11$\times$7 and 15$\times$15$\times$9  $\Gamma$-centered k-meshes are performed during the structural optimization and the self-consistent calculations. The force and energy convergence criteria are set as 0.001 eV/\AA $ $ and 10$ ^{-7}$ eV, respectively. The surface states are calculated using the slab model with thickness of 33 unit cells, realized by the OPENMX software package~\cite{new10}. The structural models are visualized via the VESTA software.

\section{3. RESULTS AND DISCUSSIONS}
After an exhaustive material screening, we find the family of A$_2$B (A = Ca, Sr, Ba; B= As, Sb, Bi) materials are potential candidates for TNLEs. To be noted, one of the A$_2$B materials, namely Sr$_2$Bi has been proposed to possess both electride and nodal line characters without considering SOC quite recently~\cite{add57}. Unfortunately, the previous work did not make detailed discussions on the SOC effect of Sr$_2$Bi. Thus, the A$_2$B materials are still an excellent choice to study the properties of TNLEs.

\begin{figure}
\includegraphics[width=10cm]{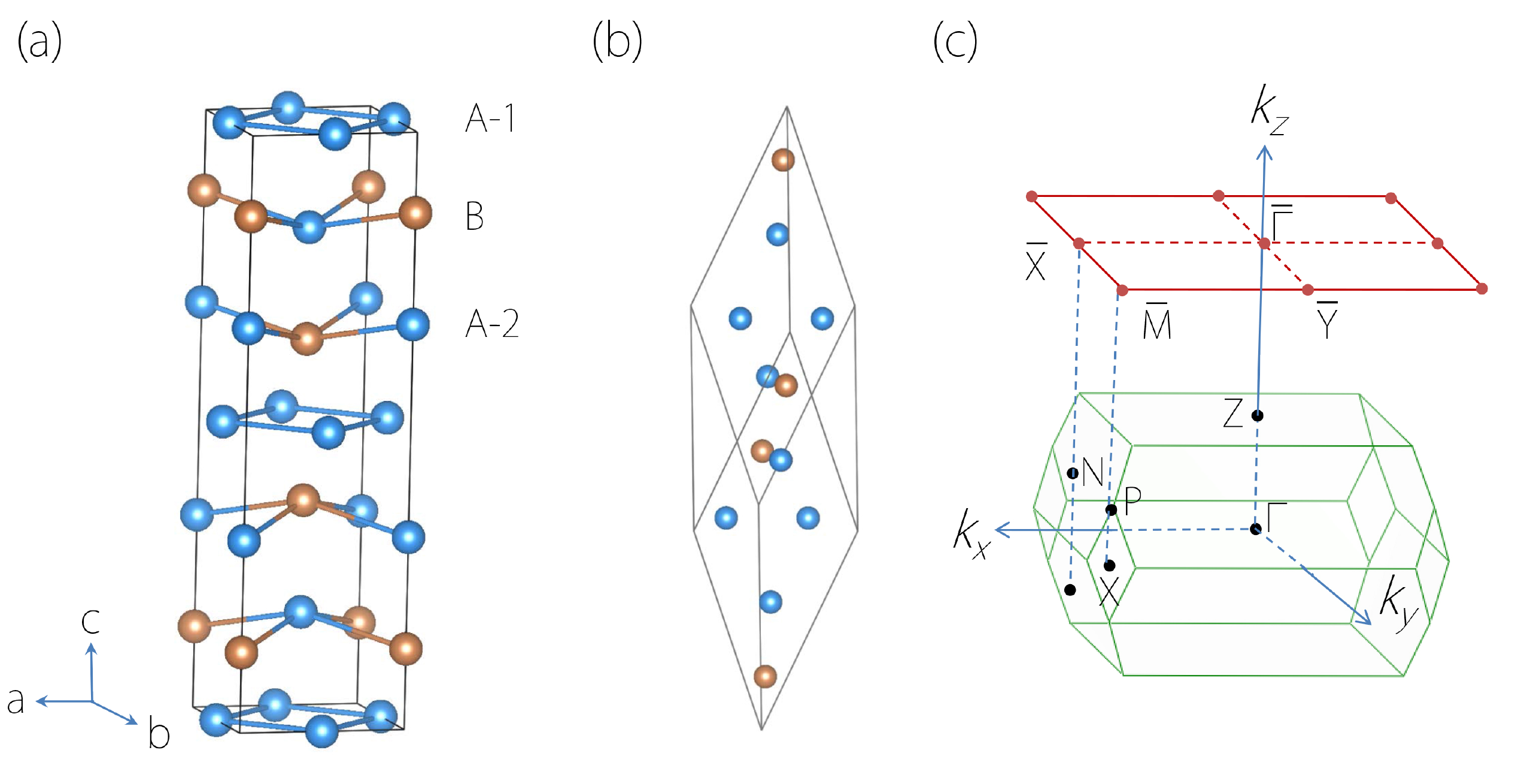}
\caption{Crystal structure of A$_2$B materials shown as (a) conventional and (b) primitive cell forms. (c) The bulk Brillouin zone and projection onto the (001) surface.
\label{fig1new}}
\end{figure}

\begin{figure}
\includegraphics[width=10cm]{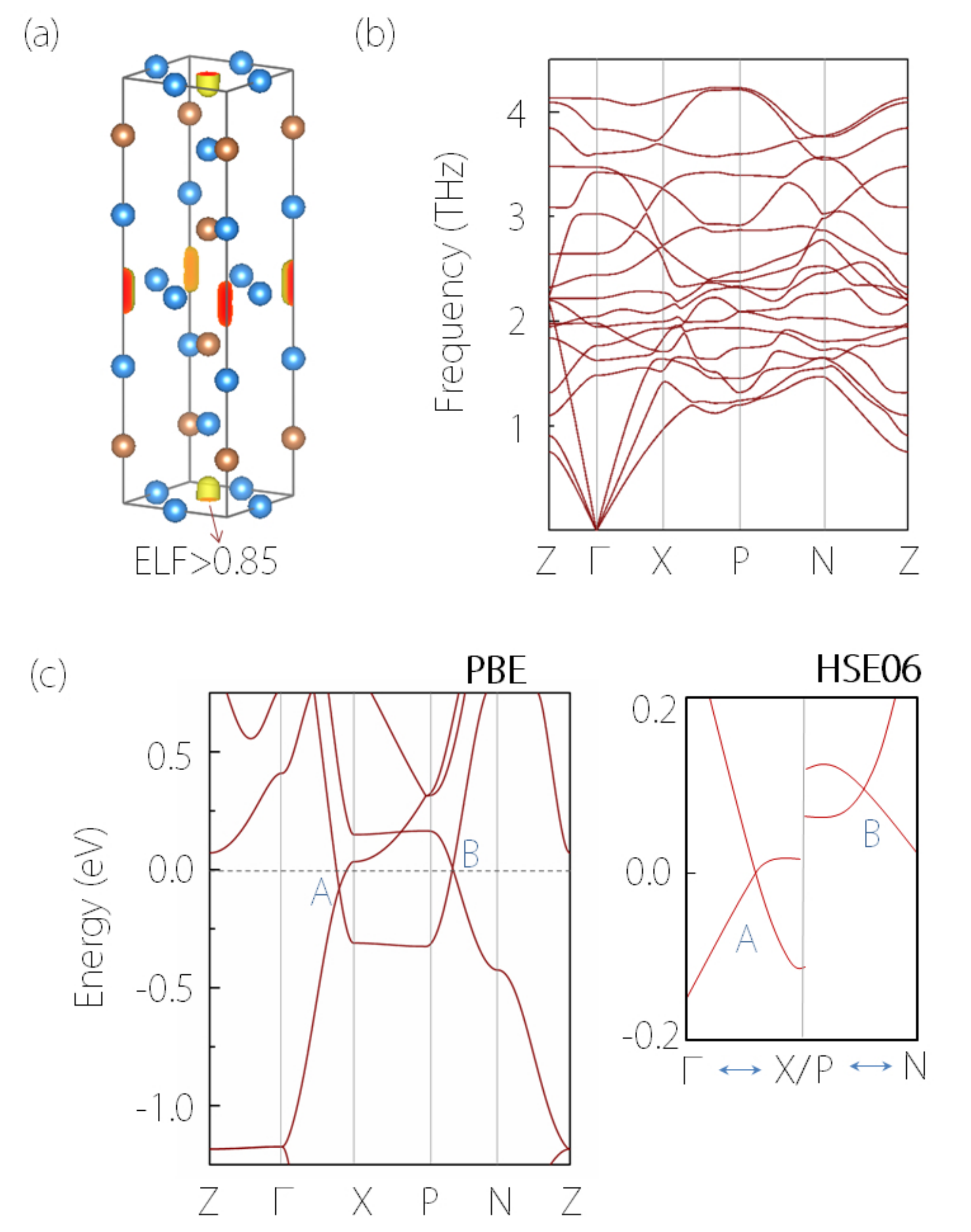}
\caption{(a) ELF graph of Sr$_2$Sb with the isosurface value of 0.85. (c) Phonon band structures of Sr$_2$Sb. (c) Electronic band structures of Sr$_2$Sb under PBE and HSE06 functional. Here the spin-orbit coupling is ignored, and the two band-crossing points near the Fermi level are labelled as A and B.
\label{fig2}}
\end{figure}

The A$_2$B compounds possess a tetragonal lattice structure with the space group \emph{I4/mmm} (No. 139). Fig.~\ref{fig1new}(a) shows the conventional cell of A$_2$B compounds, where one group of A atoms (A-1) situate at the \emph{4c} Wyckoff sites (0, 0.5, 0), and the other group of A atoms (A-2) and B atoms are at the \emph{4e} Wyckoff sites (0, 0, $\mu$/$\mu$'), respectively. One conventional cell contains two units of primitive cell, and the primitive cell representation is shown in Fig.~\ref{fig1new}(b). Moreover, these A$_2$B compounds are all existing materials that have already been synthesized experimentally~\cite{add53,add54,new1,new2,new3,new4,new5}. To further convince their stability, we calculate the phonon dispersions for the A$_2$B materials [see Fig.~\ref{fig2}(b) and Supplementary Information~\cite{add62}]. The phonon dispersions show that, these A$_2$B materials (except Ba$_2$As) have no imaginary mode throughout the BZ, indicating most of them are dynamically stable. Moreover, in A$_2$B compounds, both \emph{P} and \emph{T} symmetries preserve.


Since most of A$_2$B compounds have similar electronic structures, in the following we use Sr$_2$Sb as a concentrate example. Considering the normal charge of atoms, Sr$_2$Sb has the charge state of [Sr$_{2}$Sb]$^{+}$e$^{-}$, which contains one excess electron for per formula. The electron localization function (ELF) index shows the degree of electron localization and is proved effective to identify electrides~\cite{add55,add56,add57,add58,add59,add60}: if the target material yields large ELF values (usually $>$ 0.75) in the interstitial lattice region, it can be considered as a potential electride. The three-dimensional (3D) ELF graph of Sr$_2$Sb is shown in Fig.~\ref{fig2}(a), where we adopt the ELF isosurface value as large as 0.85. We can clearly observe that ELF values lager than 0.85 all occur in the interstitial region, indicating Sr$_2$Sb is a potential electride. The distribution of excess electrons in Sr$_2$Sb is quite similar with that in Sr$_2$Bi~\cite{add57}. Then we come to the band structure of Sr$_2$Sb. Without considering SOC, as shown in Fig.~\ref{fig2}(c), near the Fermi level there exist two band crossing points: point A at -0.079 eV along the $\Gamma$X path, and point B at 0.007 eV along the PN path. Further calculations show Sr$_2$Sb indeed undergoes band inversions near the Fermi level. To be specific, along the $\Gamma$X path, the two inverted bands have the B$_{1u}$ and A$_{1g}$ representations of the D$_{2h}$ point group at the X point; and along the PN path, they have the B$_2$ and A$_1$ representations of the D$_{2d}$ point group at the P point. Both points A and B have double degeneracy without counting spin, and they are not isolate nodal points because both \emph{P} and \emph{T} symmetries preserve in Sr$_2$Sb. Similar band structure has also been observed in Sr$_2$Bi previously~\cite{add57}. It is well known that, the PBE functional usually underestimates the size of band gap in semimetals. So we check the band structure of Sr$_2$Sb by using the hybrid HSE06 functional. From Fig.~\ref{fig2}(c), we can clearly observe that the two crossing points A and B retain under HSE06 calculation.

\begin{figure}
\includegraphics[width=10cm]{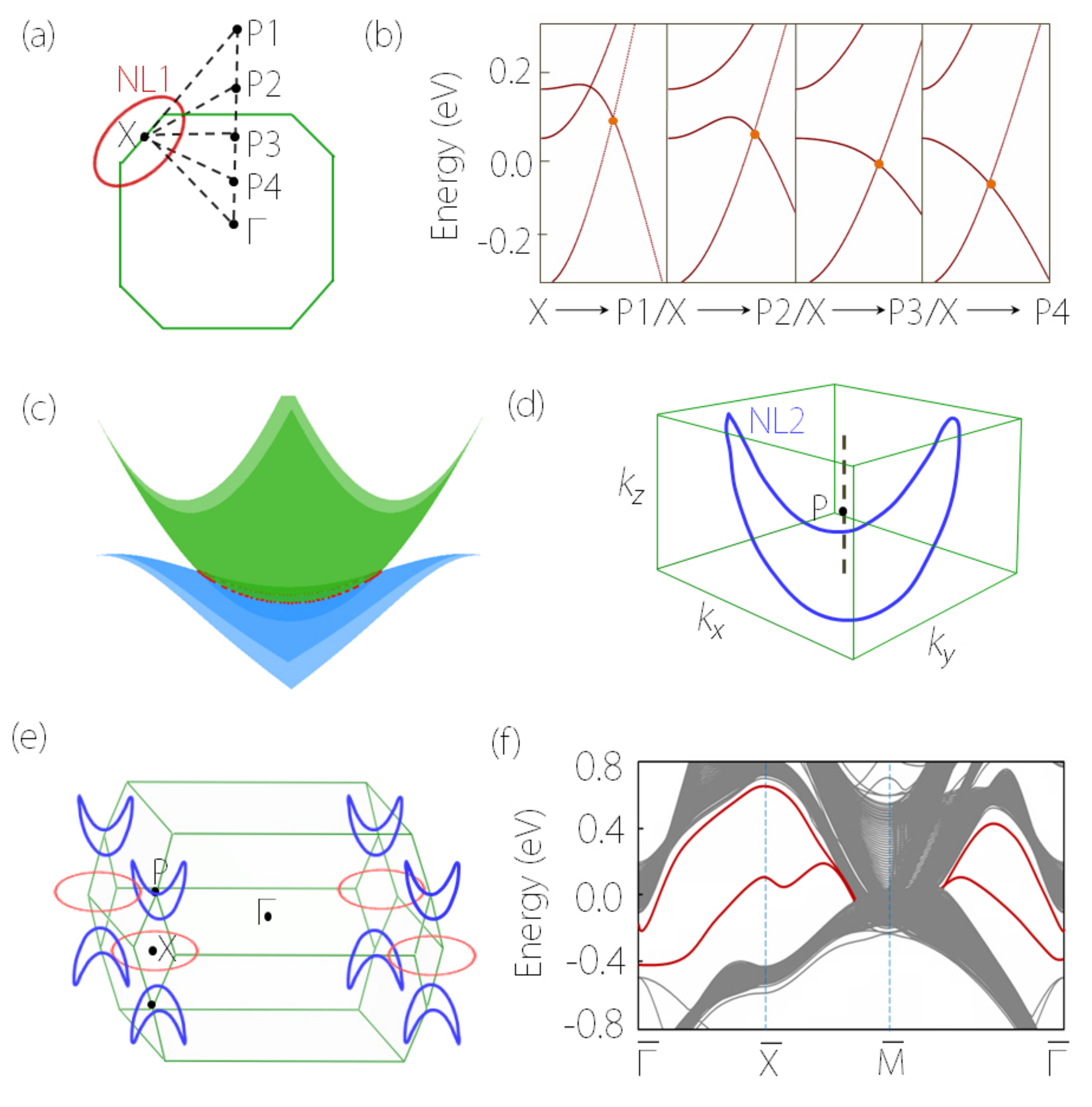}
\caption{(a) The Brillouin zone and the selected \emph{k}-paths in the \emph{k$_z$} = 0 plane. NL1 is schematically shown by the red circle. (b) Band structures along the \emph{k}-paths as shown in (a). (c) 3D plotting of the band structure near NL1. (d) Shape of NL2 in the 3D \emph{k}-space. (e) Illustration of the nodal lines in Sr$_2$Sb. (f) Band structure of Sr$_2$Sb slab with (001) surface orientation. The drumhead surface bands are highlighted by the red lines.
\label{fig3}}
\end{figure}

A careful scan of the band structure finds points A and B belong to two separated nodal lines. Point A belong to a nodal line centering the X point in the \emph{k$_x$-k$_y$} plane, which enjoys additional protection of the mirror symmetry in the plane. As shown in Fig.~\ref{fig3}(a) and~\ref{fig3}(b), we can always obtain a linear band crossing point along the \emph{k}-path starting from the X point in the plane. The shape and the band dispersion of the nodal line can be clearly shown by the 3D plotting of the band structure in the \emph{k$_z$} = 0 plane, as shown in Fig.~\ref{fig3}(c). Here we denote the nodal line as nodal line 1 (NL1) [see Fig.~\ref{fig3}(a)]. Point B belong to another nodal line (denoted as NL2) centering the P point. Different with NL1, NL2 does not reside on a specific plane but manifests a snakelike profile in the 3D Brillouin zone, as shown in Fig.~\ref{fig3}(d). As the result, Sr$_2$Sb possesses two kinds of nodal lines in the \emph{k}-space [see Fig.~\ref{fig3}(e)]. Nodal line semimetals usually characterize drumhead surface states. For Sr$_2$Sb, we show the (001) surface band structure in Fig.~\ref{fig3}(f). We can observe clear drumhead surface states originating from the nodal lines. With preserving both electride and nodal line signatures, Sr$_2$Sb is a typical TNLE. Beside Sr$_2$Sb, we find other six A$_2$B materials including Ca$_2$As, Sr$_2$As, Sr$_2$Bi, Ba$_2$As, Ba$_2$Sb, and Ba$_2$Sb in A$_2$B materials are also excellent TNLEs, while the rest Ca$_2$Sb and Ca$_2$Bi are only ordinary electrides without nodal line band structures~\cite{add62}.

\begin{figure}
\includegraphics[width=10cm]{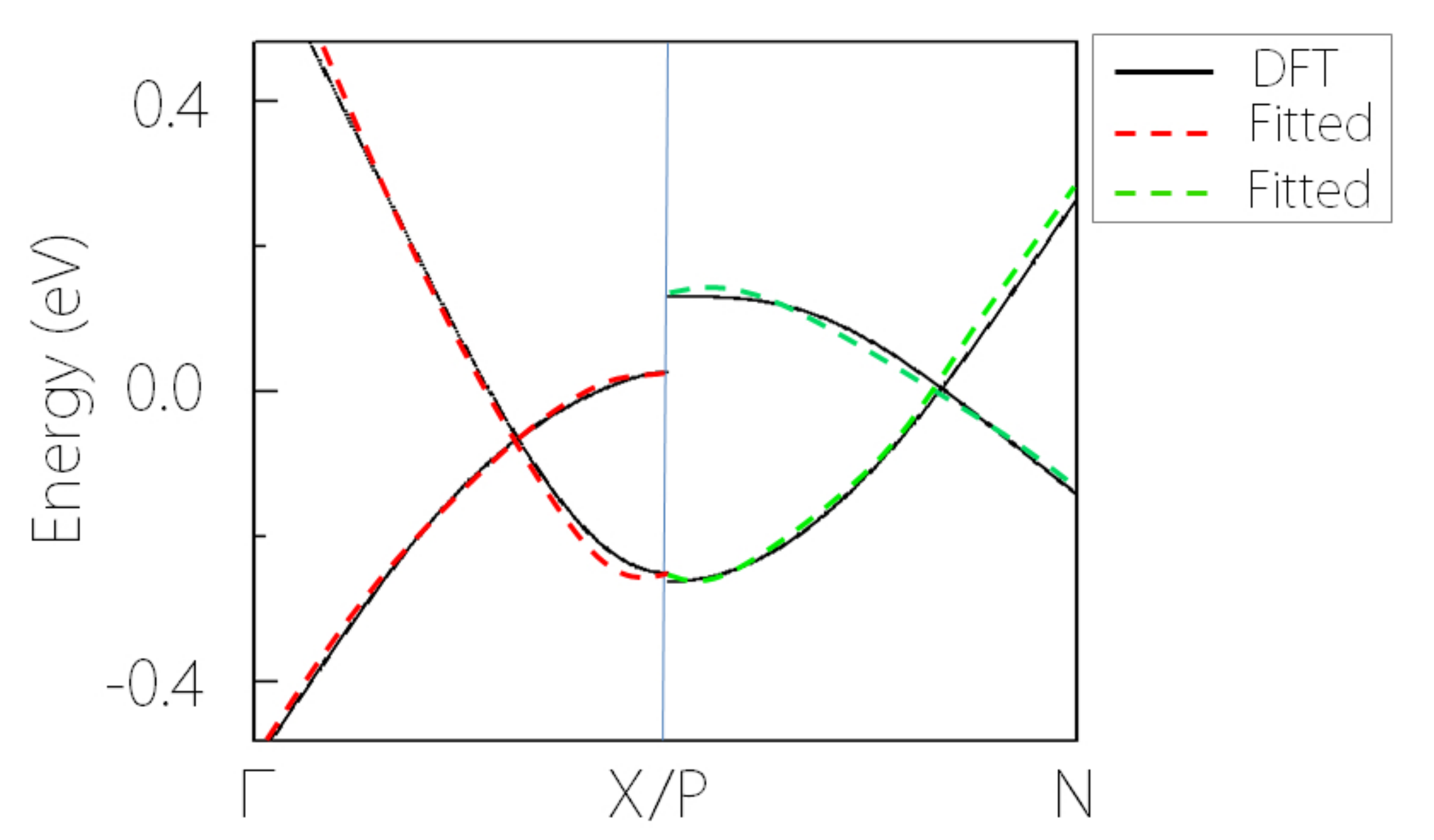}
\caption{Fitted band structures (the red/green dashed lines) using the effective model compared with the DFT band structures ( the black solid line).
\label{fig2new}}
\end{figure}

To further convince the DFT calculations, we construct effective $k\cdot p$ models to describe the nodal lines. For NL1, the crossing bands have irreducible representations  B$_{1u}$ and A$_{1g}$ at the X point. Using them as basis, we can obtain the effective Hamiltonian (up to k-quadratic order) as:

\begin{eqnarray}
H & = & \left(\begin{array}{cc}
 h_{1} & -iEk_{z}\\
iEk_{z} &  h_{2}
\end{array}\right),\label{HNR}
\end{eqnarray}
where $h_{1(2)}=d_{1(2)}+a_{1(2)}k_{x}^{2}+b_{1(2)}k_{y}^{2}+c_{1(2)}k_{z}^{2}$. For NL2, the crossing bands possess irreducible representations B$_2$ and A$_1$ at the P point. And the effective Hamiltonian yields to be:
\begin{eqnarray}
H & = & \left(\begin{array}{cc}
 h_{3} & -i(Fk_{x}k_{y}+Gk_{z})\\
i(Fk_{x}k_{y}+Gk_{z}) &  h_{4}
\end{array}\right),\label{HNR}
\end{eqnarray}
where $h_{3(4)}=d_{3(4)}+a_{3(4)}(k_{x}^{2}+k_{y}^{2})+c_{3(4)}k_{z}^{2}$. In above Hamiltonian, the parameters including $a_{1(2,3,4)}$, $b_{1(2)}$, $c_{1(2,3,4)}$, $d_{1(2,3,4)}$, \emph{E}, \emph{F}, and \emph{G} are material-specific. The fitted band structure from the modal and the DFT results are shown in Fig.~\ref{fig2new}, which exhibit a good agreement.

\begin{figure}
\includegraphics[width=10cm]{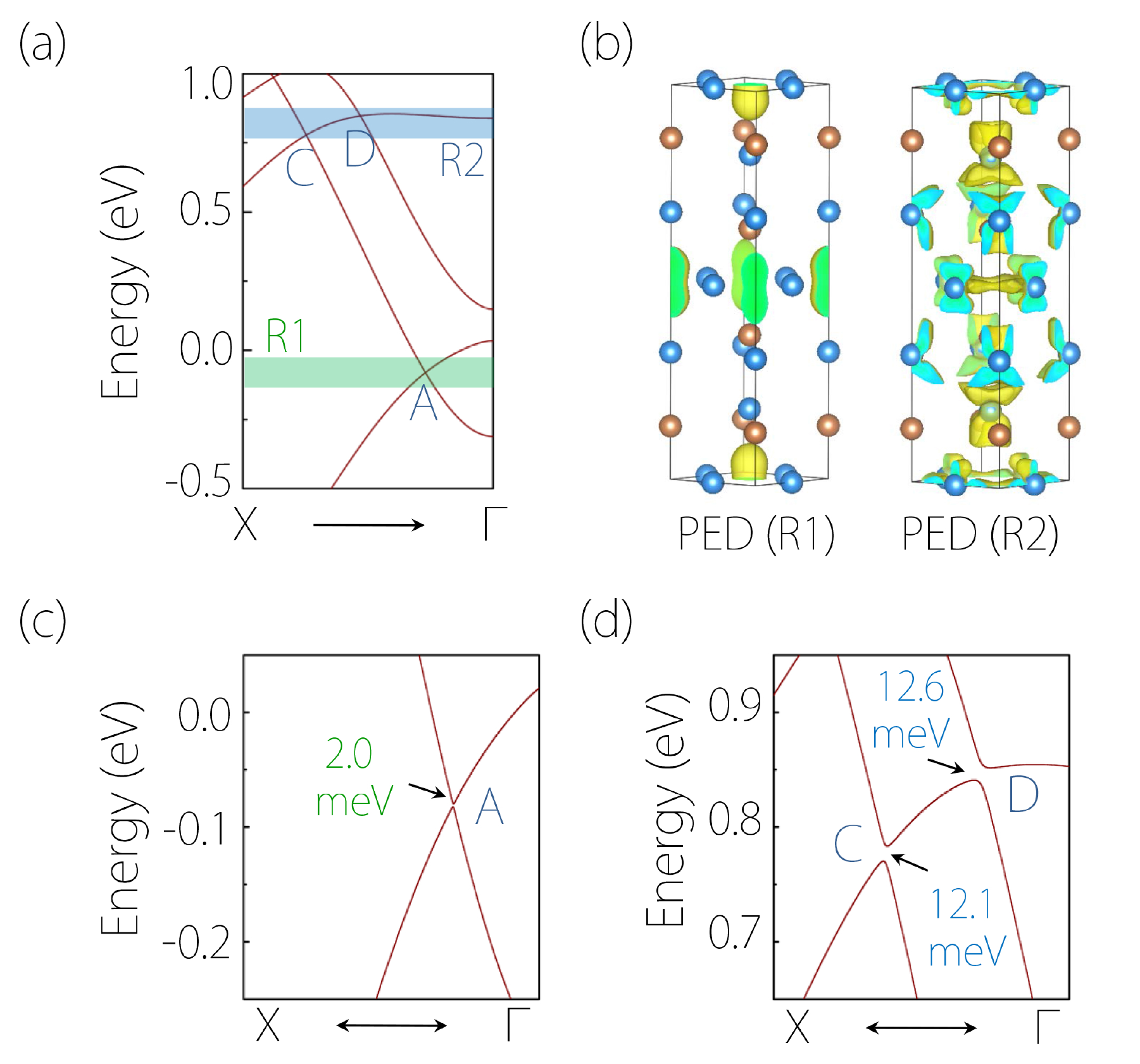}
\caption{(a) Enlarged band structures along the X$\Gamma$ path of Sr$_2$Sb without SOC. The three band crossing points are denoted as A, C, and D, respectively. An energy region of 0.1 eV around crossing point A ($|$E -E$_{A}$$|$$<$0.05 eV) is denoted as R1, and that around crossing points C and D is denoted as R2. (b) The PED graphs for the electronic states of R1 and R2. The isosurface value is chosen as 0.00065 Bohr$^{-3}$ for R1 and 0.00085 Bohr$^{-3}$ for R2, respectively. (c) Band structures around crossing point A with SOC included. (d) is similar with (c) but for the band structures around crossing points C and D. In (c) and (d), the size of SOC-induced gap is indicated.
\label{fig4}}
\end{figure}

As mentioned above, the nodal lines in Sr$_2$Sb are protected by the coexistence of \emph{P} and \emph{T} symmetries, and they will be gapped when SOC is taken into account. However, for electride Sr$_2$Sb, the nodal lines are mostly contributed by the non-nucleus-constrained electrons, thus the SOC-induced gaps are expected to quite small although both Sr and Sb are heavy elements. This insight has been verified by our computations. In Fig.~\ref{fig4}(a), we show the enlarged SOC-absence band structure of Sr$_2$Sb along the X$\Gamma$ path, which includes crossing point A of NL1 (see the shadowed region R1). We calculate the partial electron density (PED) of R1 (-0.129 eV to -0.029 eV) to identify the electron distributions around crossing point A. As shown in the left panel of Fig.~\ref{fig4}(b), we find the electrons for R1 origin from interstitial electrons. After including SOC, the band structure of crossing point A is indeed nearly unaffected, where the SOC-induced gap is negligible [2.0 meV, see Fig.~\ref{fig4}(c)]. Beside crossing point A, we notice their exist other two band crossing points [points C and D in the shadowed region R2, see Fig.~\ref{fig4}(a)]. Being Different with R1, the band structure in R2 is no longer contributed by interstitial electrons but by traditional bonding electrons, as shown by the PED graph for R2 [see the right panel of Fig.~\ref{fig4}(b)]. As a result, the SOC-induced gaps at crossing points C and D are several times larger than that at point A [see Fig.~\ref{fig4}(c) and~\ref{fig4}(d)]. This indicates the size of SOC-induced gap is quite relate to the distribution of electrons. In TNLE Sr$_2$Sb, the nodal lines are almost contributed by the interstitial electrons and are nearly unaffected when SOC is included.

Then we make a comparison between present TNLEs A$_2$B compounds and existing typical NLSs. As shown in Fig.~\ref{fig1}, the TNLEs in A$_2$B materials have very small SOC-induced gaps and all situate far below the region of 0.5$<$R$<$2.0, indicating nodal lines in TNLEs have significantly smaller SOC-induced gaps comparing with traditional NLSs. Therefore, TNLEs are indeed excellent NLSs with pronounced nodal line character. It should be noted that, in TNLEs the SOC-induced gaps at different parts of the nodal line may be different because of the hybridization with orbitals of the interstitial bands. Beside the band structure at $\Gamma$X path, we have also checked the SOC effect at other parts of the nodal line. Our calculations show that the SOC-induced gap does not change much at different \emph{k}-paths in these A$_2$B materials. To be specific, the SOC-induced gaps along the nodal line vary in 2.6 meV in Sr$_2$Sb, 1.8 meV in Ca$_2$As, 3.2 meV in Sr$_2$As, 9.8 meV in Sr$_2$Bi, 4.6 meV in Ba$_2$As, 3.9 meV in Ba$_2$Sb, 6.7 meV in Ba$_2$Bi, respectively. Moreover, we also find the SOC-induced gaps can be tailored by tuning the distribution or localization of electrons around the nodal lines in TNLEs~\cite{add62}. The controllable SOC-induced gaps make TNLEs further practical to realize the nodal line states.

Before closing, we would like to emphasize that, the main focus of our work is quite different from previous works. In particular, in Ref.[~\cite{add57}], Hirayama \emph{et al.} proposed that electrides can also obtain band inversions which are needed for topological materials including topological insulating and topological semimetal phases. They also gave out several candidate materials (including Sr$_2$Bi) for these topological phases in electrides. However, they neither made detailed study on the electronic structures of topological electrides under SOC, nor focused on topic of TNLEs for realizing small SOC-induced gaps in nodal lines.

\section{4. SUMMARY}
In summary, by promoting the realization of TNLEs in A$_2$B (A = Ca, Sr, Ba; B= As, Sb, Bi) materials, we provide new insight on designing excellent NLSs with the nodal line structure nearly unaffected by the SOC effect. TNLEs possess both the electronic signatures of nodal lines and electrides. Especially, in TNLEs the nodal lines are mostly contributed by interstitial electrons, which are nearly unconstrained by the nuclei, thus SOC has weak impact on nodal lines even containing heavy elements. Besides, unlike traditional NLSs, we find the SOC-induced gaps in TNLEs are tunable by external perturbations such as lattice strain. TNLEs provide excellent platform to realize the exotic properties from the intrinsic nodal line state.

\section{Conflicts of interest}
There are no conflicts to declare.

\section{Acknowledgements}
The authors thank Shan Guan and Zhiming Yu for helpful discussions. This work is supported by National Natural Science Foundation of China (Grants No. 11904074), Natural Science Foundation of Hebei Province (Grants No. E2019202222 and No. E2016202383) and the National Key R\&D Program of China (Grants No.  2016YFA0300600), the NSF of China (Grants No. 11734003), and the Project of Scientific Research for High Level Talent in Colleges and Universities of Hebei Province (No. GCC2014042). One of the authors (X.M. Zhang) acknowledges the financial support from Young Elite Scientists Sponsorship Program by Tianjin. One of the author (G.D. Liu) acknowledges the financial support from Hebei Province Program for Top Young Talents. One of the authors (B. Fu) acknowledge Sichuan Normal University for financial support (No. 341829001).

\end{document}